# Laser spectroscopy of finite size and covering effects in magnetite nanoparticles

V. N. Nikiforov[1], A.N. Ignatenko [2], A.V. Ivanov[3*] and V. Yu. Irkhin[2]

[1] M.V. Lomonosov Moscow State University, Moscow
[2] Institute of Metal Physics, Ural Division of RAS, Ekaterinburg
[3] N.N. Blokhin Russian Cancer Research Center, Moscow
*E-mail: ivavi@yandex.ru

**Abstract.** The experiments on the impact of the size of magnetite clusters on various magnetic properties (magnetic moment, Curie temperature, blocking temperature etc.) have been carried out. The methods of magnetic separation, centrifuging of water suspensions of biocompatible iron oxide nanoparticles (NPs) allow producing fractions with diameter of nanoparticles in the range of 4÷22 nm. The size of NPs are controlled by the methods of dynamic light scattering (DLS), transmission electron microscopy (TEM) and atomic force microscopy (AFM). For the first time the DLS method is applied in real time to control the size during the process of the separation of the NPs in aqueous suspensions. The changes of the size of NPs cause a shift in the Curie temperature and in the changes in the specific magnetic properties of the iron NPs. The experimental data is interpreted on the basis of Monte Carlo simulations for the classical Heisenberg model with different bulk and surface magnetic moments. It is demonstrated experimentally and by theoretical modeling that magnetic properties of magnetite NPs are determined not only by their sizes, but also by the their surface spin states, while both growing and falling dependences of the magnetic moment (per $Fe_3O_4$ formula unit) being possible, depending on the number of magnetic atoms in the nanoparticle. Both NPs clean and covered with a bioresorbable layer clusters have been investigated.

PACS: 42.62.Fi, 75.75.-c, 36.40.Cg

*Key words: dynamic light* scattering *spectroscopy, magnetic nanoparticles; magnetite; magnetization; Monte Carlo simulations*

## 1. Introduction

The method of Laser correlation spectroscopy namely dynamic laser scattering (DLS) is used in nanotechnology for reliable control of the size of nanoparticles (NPs) in real time, including the actual chemical and biological processes [1-3]. Nanoscale magnetic materials are of particular interest for applications in ferrofluids, high-density magnetic storage, high-frequency electronics, high-performance permanent magnets, magnetic refrigerants. Magnetism of NPs is the area of intensive development which impacts various areas of research including material science, condensed matter physics, biology, medicine, biotechnology, planetary science, and so on [4–6]. In particular, iron oxide colloids have a low toxicity and manifest good biocompatibility, which makes them applicable for various areas of medicine, e.g. for drug delivery systems and for hyperthermia treatment of cancer. For the use in magnetic separation, MR tomography, magnetic hyperthermia and other applications [7], the methods of metrological control of magnetic NPs are developed [8]. Today this is quite a difficult task, since in the 1÷10 nm size range many techniques used are at the limit of their resolution, and the data obtained by different methods do not always correlate. For the drug development, the most difficult question is the control of the size of the particles and, in the first place, the size of these objects in native fluids. The key techniques for the creation of the composite materials to be used in oncology are based on the fact that typical range for their size allows them to accumulate selectively in malignant tumors as compared to healthy tissue, in which the affected area is irradiated in a certain spectral region [7]. So, the dispersion of nanoshells should be strictly limited to a narrow region.

The increasing interest in nanoobjects is associated with the manifestation of the so-called "quantum size effects". These effects take place with the decrease of the size of the particle and with the transition from a sample of macroscopic size to the sample size of a few hundred or thousand of atoms, when the electron spectrum in the valence and conduction bands change sharply. This phenomenon affects electron and magnetic properties of the particles. The continuous spectrum presented at the macro-scale is replaced by a set of discrete levels, the distances between which are dependent on the size of the particles. Due to such a size-dependency of the NPs, they are used for new type of applications.

At present, it has been firmly confirmed by various studies that nanostructures (including nanoclusters) demonstrate a significant difference in many physical and physical–chemical properties in comparison with bulk materials [9, 10]. For example, nanoclusters melting temperatures could be both above and below those of their bulk analogues [11-13].

As regard for the works in the field of physics of magnetic nanoclusters (see e.g. [14]), a number of models have been constructed and many experiments have been performed. However, systematical analysis of the conducted experiments has not yet been performed. Now, it is possible to believe with a large extent that the study of NPs leads to necessity of accepting two new arguments [9, 10]:

1. Elementary excitation spectrum of NPs is discrete owing to their small size.
2. The total number of surface states of NP is comparable to the number of the bulk ones.

The use of these two principles concerning "nanomagnetism" can be realized by means of "core-shell" picture, which allows separating the contributions of the surface and bulk states to the magnetic properties of a NP. In this paper these ideas are applied to iron-oxide (magnetite) NPs. The models proposed earlier [8] include a phenomenological Weizsäcker model (by analogy with the Weizsäcker approach in nuclear physics) and microscopic Ising and Heisenberg models with modified surface magnetic moments. Our approach is based on both making experiments and constructing models. The experimental research deals with magnetite NPs synthesized by different methods. DLS technique, supposedly, should have simple and effective scheme feedback, automatically adapting the power of the laser source and under study of a particular sample; improved characteristics of the receiving path for the input signal with a low noise amplifier and squelch; efficient algorithm for decomposing the spectrum of the scattered signal power fluctuations into the components that characterize the individual ("mono-disperse") fraction of NPs presented in the system; this algorithm should include the possibility of simultaneous resolution of the integral equations describing the scattered signal for different scattering angles and regularization methods. Dynamic light scattering allows controlling the distribution of the size of inorganic iron oxide NPs in suspension. Apparatus based on the DLS method allows controlling the production of nanomaterial at all stages of the technological cycle.

## 2. Materials and experimental methods

Native $Fe_3O_4$ was synthesized according to the following method: firstly, particle sedimentation from iron salt solution mixed with natrium hydroxide at pH 10÷12, secondly, activation in dilute HCl, thirdly, stabilization by treatment with dextran, and, finally, purification [15, 16].

The iron oxide NPs were synthesized by co-precipitation [17]. To extend the region of SPION (superparamagnetic iron oxide NPs) samples sizes, the conditions of the synthesis were changed with careful control by transmission electron microscopy (TEM) and X-ray diffraction, while low-quality samples were eliminated [7]. SPION samples were coated by DNA [18] and by PVA [19, 20]. The NPs were stabilized *in situ* by organic surfactant molecules which acted both as a stabilizer of the microemulsion system and as a capping layer of the NP surface. The control of the NP size was attained by careful adjustment of the preparation conditions. The structure, morphology and the distribution of NPs size were investigated by DLS, X-ray diffraction, TEM, atomic force microscopy (AFM) and scanning electron microscopy.

Physical and chemical characterization of initial precipitates, $Fe_3O_4$, dextran-ferrite and their suspensions were performed by DLS laser technique KURS3M (Fig.1), TEM, electron spin-resonance spectroscopy (ESR), magnetic measurements (by vibrating and SQUID magnetometers) and X-ray diffraction.

The diameter of magnetic NPs is the main parameter, determining their properties, such as toxicity, biocompatibility, physical and chemical properties [7]. Nevertheless, the covering of NPs can also affect their magnetic properties. The DLS technique helps detecting, in real time, the changes of the size of NPs that, consequently, leads to a shift of the Curie temperature and specific magnetic properties of iron NPs. Magnetic separation [7] and centrifuging of water suspension of biocompatible iron oxide NPs allow producing "monodisperse" fractions of NPs (Fig.1, left) with different diameters (4÷22 nm) controlled by the methods of dynamic light scattering, X-ray diffraction, TEM, atomic force microscopy (AFM) and scanning electron microscopy.

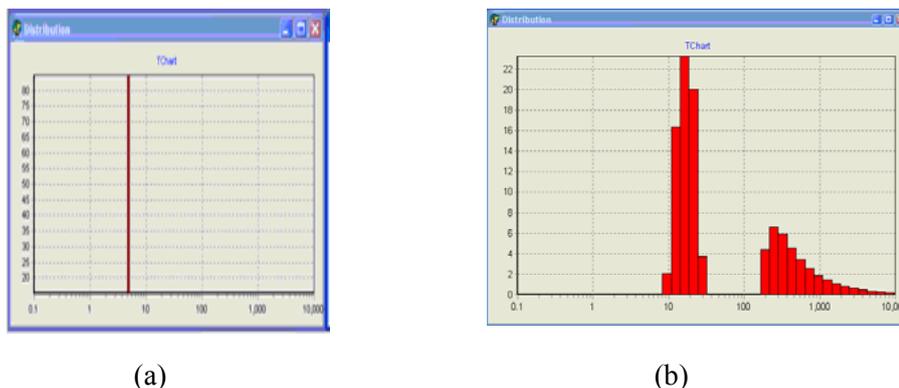

(a)            (b)

**Figure 1.** Magnetic separation, centrifuging of water suspension of biocompatible iron oxide NPs produce fractions with different diameters of NPs 4-22 nm controlled by dynamic light scattering technique. The "monodisperse" fraction (a) is shown on the left, in the right the source DLS spectrum (b) before treatment is presented.

The TEM was performed using EM-400 instrument (500.000x, a resolution of 0.4 nm) and a SEM IPS image analyzer. Crystal structure was depicted, and composition was controlled to avoid multiphase samples. Test results confirm the presence of one phase of the spinel structure, as well as the size of an average particle of 4÷22 nm (by Scherrer equation).

A particular effort was made to study the effect of the size and capping of the NPs on their magnetic structure. Magnetic diagnostics methods SQUID and ESR [21] were used. The temperature and magnetic field dependences on magnetic moment were measured, and ESR spectra for NPs were obtained at temperatures of 4.2÷380 K. Specific saturation magnetization $J_s$ of powder samples and their temperature dependence at 10–700 K was determined on a vibrating-sample magnetometer with a sensitivity of $10^{-5}$ A·m$^2$/kg in the applied magnetic field about 0.75 T. ESR spectra in 78–400 K temperature interval were collected. Temperature and magnetic field dependences also have been measured at 4.2÷350 K by SQUID and vibrating sample magnetometer VSM PARC-155 as well as by Curie balance magnetometer.

## 3. Results of experiments

The distribution of magnetic NPs in size and determined their average size were obtained experimentally. The co-precipitation method enabled to obtain the spectrum from 5 to 22 nm. The application of centrifugation, magnetic separation, and shaking with ultrasound processing yielded fractions with specific dimensions.

Besides that, coated iron oxide NPs in the polymer matrix have been investigated. According to TEM, AFM, DLS their size was 60÷180 nm. Electron microscopy (SEM) demonstrated larger sized (up to 225 nm). It should be noted that the true diameter of the magnetic NPs can be smaller by several times because of passivating coating effect.

The average values of effective magnetic moment per formula unit, the effective magnetic moment of one particle, $\mu(N)$, and the average number of formula units in one particle were estimated, as in [17, 21] from the Langevin expression $M(H) \propto \mu(N)\left[\text{cth}(\mu(N)H/k_BT) - k_BT/(\mu(N)H)\right]$.

The experimental results can be summarized as follows:
1. Specific magnetic moment of nanoclusters changes monotonically with the increase of a particle size $\mu(N)$ (see figure 1).
2. This monotonic dependence of $\mu(N)$ can be either increasing (see figure 1) or falling (see figure 2) for different samples.
3. In dependences *M(H)*, the absence of hysteresis was observed, which indicates the absence of the coercive force and, consequently, the superparamagnetic state of NPs (see figure 4).
4. Curie temperature gradually falls with decreasing of the NPs size.

The dependence of the average magnetic moment change per formula unit in NPs was obtained. In our case, it was falling (see figure 2). However, in some cases one can observe the dependence of the opposite type. With the increase in the diameter of iron oxide NPs subjected to surface treatment and sealed in a polymer matrix, the rise of the magnetic moment is observed with increasing diameter of the NPs (see figure 3).

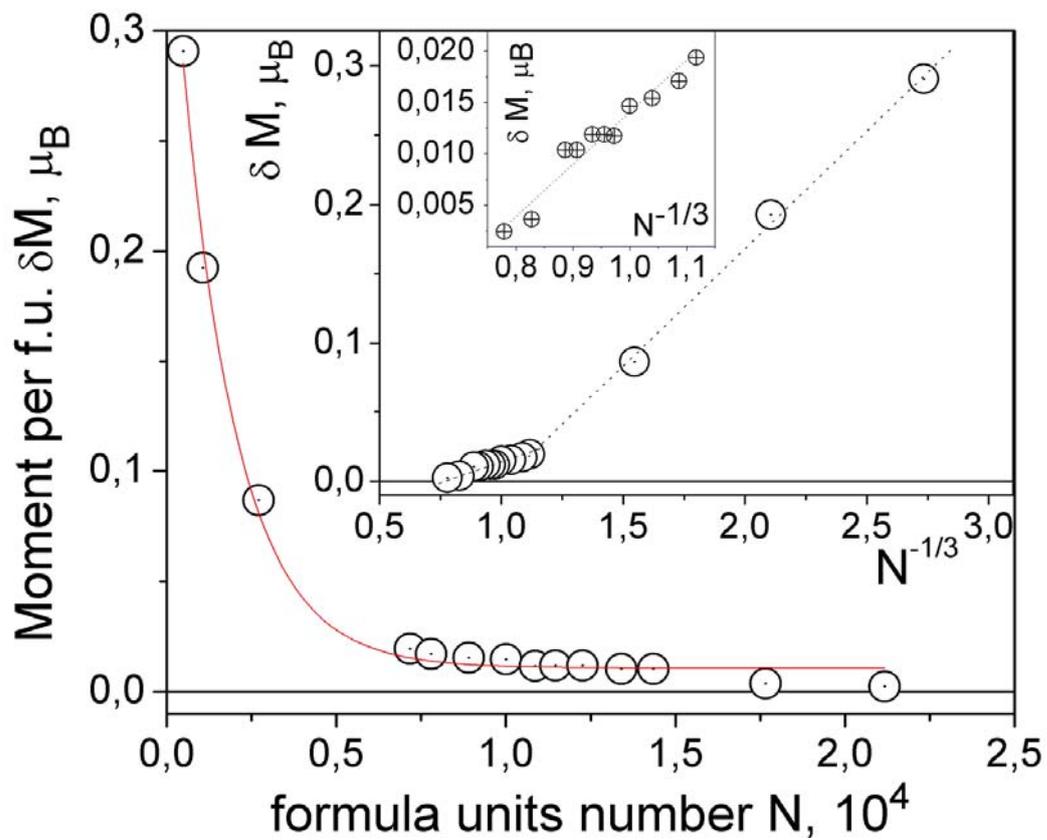

**Figure 2.** Deviation of magnetic moment per formula unit from the bulk value for magnetite 4÷22 nm NPs (in Bohr magnetons) from the Langevin formula, T = 300 K, measurement field 0.5 T. The Inset demonstrates size dependence (diameter of NPs is proportional to $N^{1/3}$) of magnetic moment per formula unit for ultra-small (see in right part of Inset) and more large NPs (see left part of Inset, and Inset in Inset too).

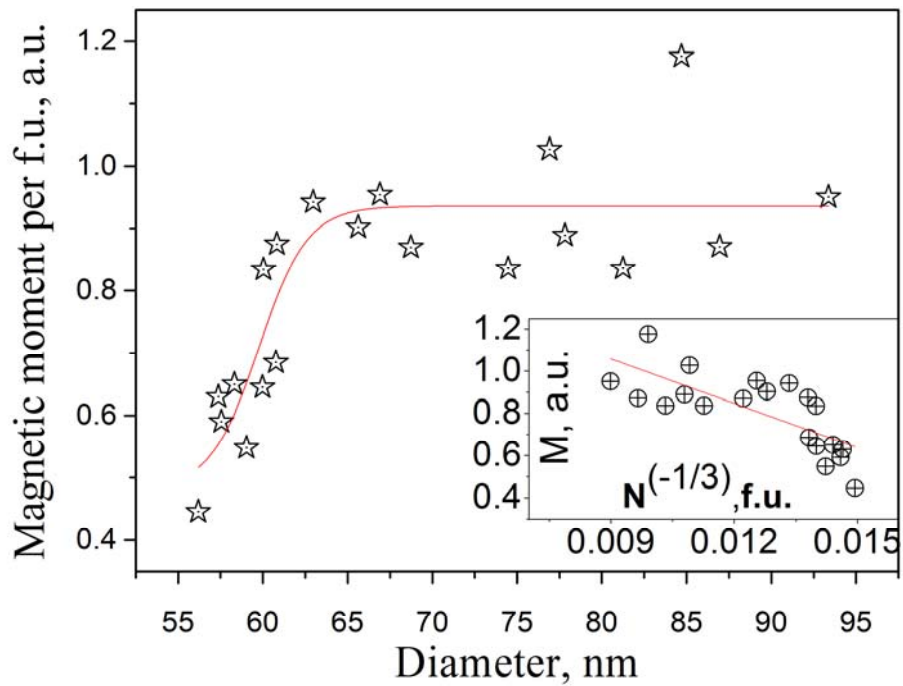

**Figure 3.** Magnetic moment per formula unit (in arbitrary units) versus diameter of magnetite NP in coated non-native iron oxide clusters in polymer matrix (see below in figure 4), T=300K. The Inset shows magnetic moment vs $N^{-1/3}$ dependence. N is the number of formula units per NP.

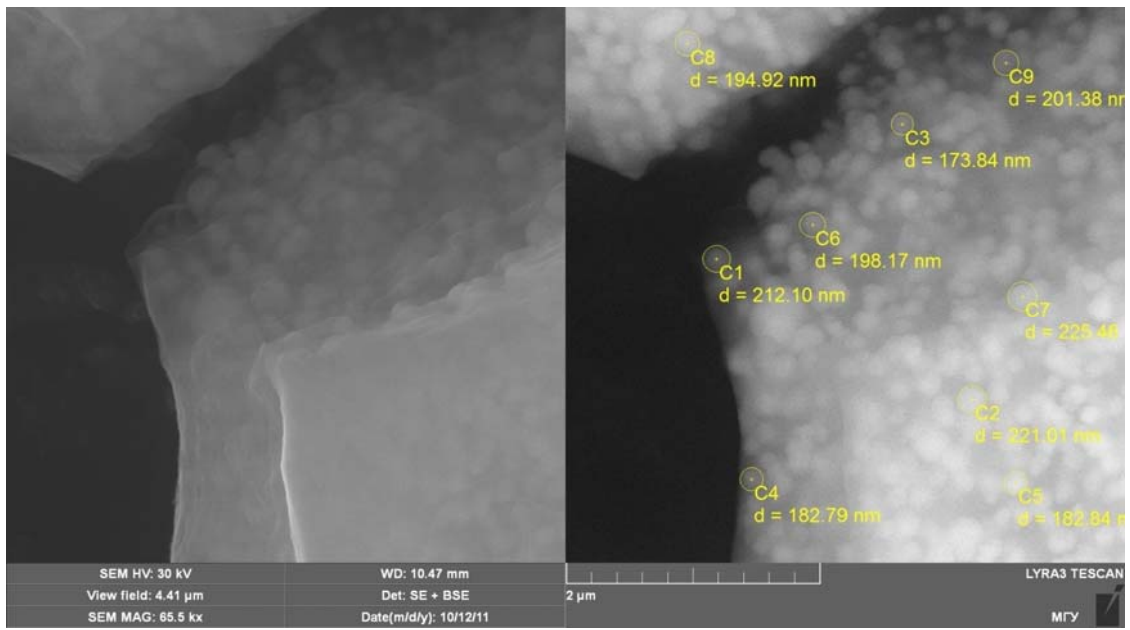

**Figure 4.** SEM image of the iron oxide NPs in polymer matrix

In the results presented here, the samples had the form of nanopowder. Moreover, a number of investigations of NPs in liquid were carried out.

A typical *M-H* (magnetization versus applied magnetic field) behavior of magnetic NPs in water is presented in Figure 5. The calculated value of magnetic moment of each cluster (according to the Langevin equation) is 6885 Bohr magnetons. The dependence with zero coercive force is a characteristic for superparamagnetic NPs.

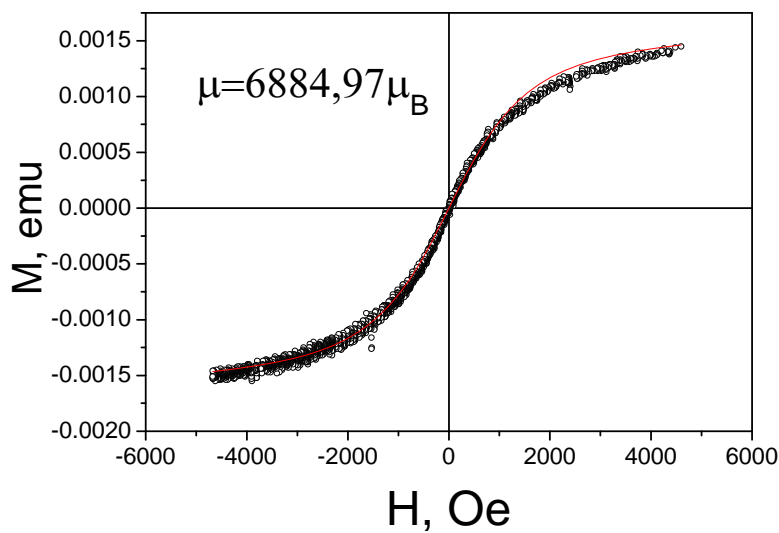

**Figure 5.** Magnetization versus applied magnetic field for a dextran-coated iron-oxide NPs in water. The solid line shows the Langevin behavior

Magnetic characteristics of NPs on the magnetite basis in the form of magnetic liquid, polymer matrix, and quarts matrix were also investigated.

The temperature dependence of the ZFC and FC magnetization in a weak field 0f 2.27 mT is shown in figure 6. Note that the corresponding values of zero-temperature magnetization $M_0$ are small. They increase rapidly with increasing of the field (see, e.g., Fig.2 in Ref. 20). In the case of bulk samples of magnetite, the Curie temperature is 580º C, as confirmed by our experiments (Fig.6 bulk curve). The temperature dependence of the magnetization of bulk samples from natural Urals magnetite and magnetite NPs with various sizes are shown in Figure 6.

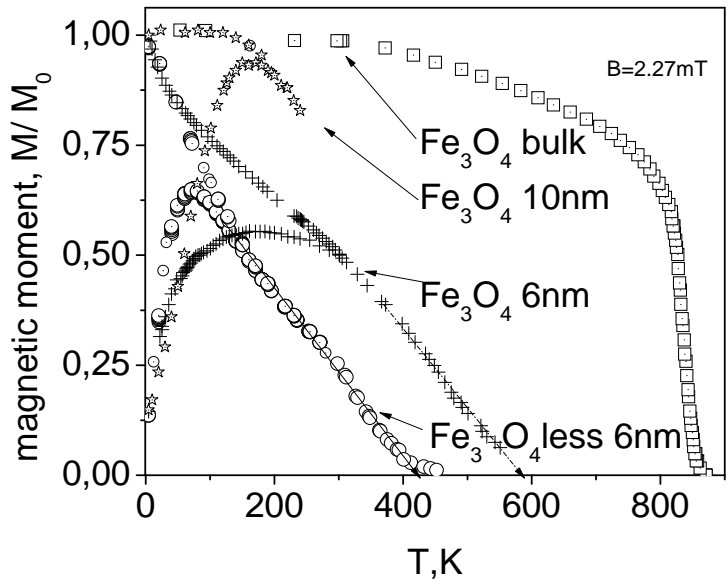

**Figure 6.** The temperature dependence of the relative ZFC/FC magnetization for bulk samples of magnetite ($M_0$=2.1 A m$^2$/kg, squares) and for dextran-coated NPs with the sizes 10 nm ($M_0$=0.86 A m$^2$/kg, asterisks), 6 nm ($M_0$ = 0.83 A m$^2$/kg, crosses), and smaller than 6 nm ($M_0$ = 0.76 A m$^2$/kg, circles).

The influence of NP environment and conditions of the synthesis on the blocking temperature $T_B$ (which corresponds to the maximum of the ZFC curve) were investigated. The dependence of $T_B$ on the synthesis conditions was revealed. The SQUID magnetic diagnostic methods were used. The magnetic properties of NPs obtained by identical methods were studied by measuring the temperature dependence of magnetization $M(T)$ and hysteresis in the $M(H)$ curves. The temperature dependence of the magnetization $M(T)$ in ZFC mode (cooling in zero field and measuring with subsequent heating) and FC mode (the same in non-zero applied field) differ significantly (see figure 6).

Since $T_B$ depends on the size of the magnetic NPs, it is promising to use this parameter for the metrology of magnetic NPs. $T_B$ is usually related to the particle size while the constant of magneto-crystalline anisotropy $K$ is defined by the equation $K = 25k_B T_B/V$, where $k_B$ is the Boltzmann constant, $V$ the average volume of one NP. However, according to our experiments, the size of magnetic NPs is not the only parameter determining the value of the blocking temperature (see Table 1).

**Table 1.** Effect of surfactants on the blocking temperature $T_B$ in iron oxide NPs, where $C_v$ - NPs concentration.

| $Fe_3O_4$ | $T_B$ |
|---|---|
| $Fe_3O_4$ PVS, $C_v$ = 1% | 70.3 K |
| $Fe_3O_4$ PVS, $C_v$ = 1%, H=0 | 90 K |
| $Fe_3O_4$ liquid col1 DNA | 61.5 K |
| $Fe_3O_4$ polymer matrix | 177 K |
| $Fe_3O_4$ glass | 145 K |

The passivation and drying of the NPs in zero magnetic field ($H$ = 0) increases the value of the blocking temperature $T_B$ significantly (up to 20 K). The coating of magnetic NPs by biopolymers, in particular by DNA, is the factor that changes the surface contribution to the magnetic moment and, according to our experimental data, has a significant impact on $T_B$ (see table 1). Coating of magnetic NPs by PVS and glass at various modes of passivation also affects the $T_B$ value.

In the case of magnetic magnetite NPs the situation changes, namely, the Curie temperature decreases with the decreasing size of NPs (see figure 7). The monotonic and size dependent Curie temperature is observed in native (non-coated) iron oxide NPs. The change of the synthesis conditions and size separation allow making variations in the diameter of magnetic NPs.

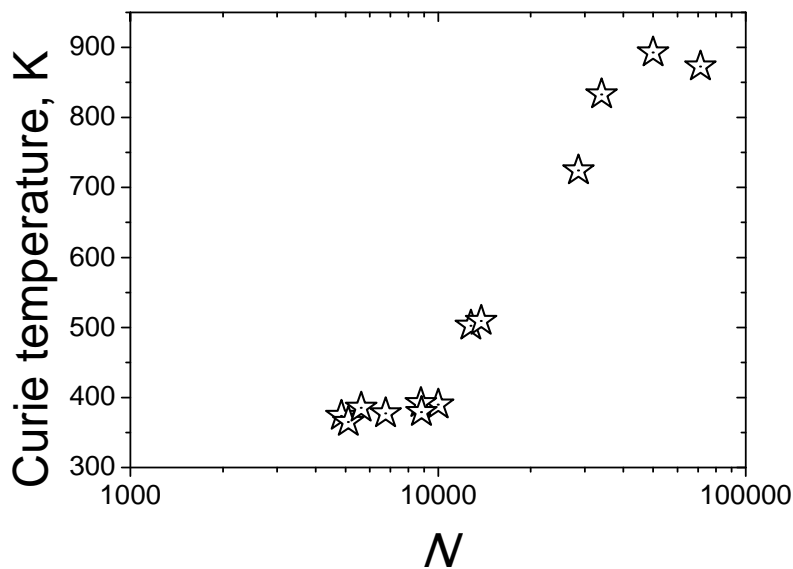

**Figure 7.** The size-dependent Curie temperature (in Kelvins) for the synthesized magnetite NPs. $N$ is the number of formula units in each magnetic iron-oxide NP.

## 4. Theory: core-shell model and Monte Carlo calculations

The experimental data presented can be interpreted on the basis of several different methods: various approximations in the Ising or Heisenberg models, quantum-chemical and first-principle calculations [22]. In the variety of these approaches, the most suitable approximation is regarded the core-shell model [5] which is a modification of the Weizsäcker model used for the analysis of specific characteristics of complex nucleus. This model covers different bulk and surface magnetic moments (per atom) for a NP [8]. Some *ab initio* band calculations predict the enhancement of the magnetic moments in thin films as compared to the bulk value [23], and it's believed that the same outcome for the surface layers of NPs could be anticipated. However, the reduction of the saturation of magnetization $M_S$ is a common experimental observation in many fine-particle systems [24]. In early models, this fact was interpreted by postulating the existence of a passive (i.e. "dead") magnetic layer originated from the demagnetization of the surface spins, which causes a reduction of $M_S$ [25]. A random canting of the surface spins caused by competing antiferromagnetic interactions between sublattices was proposed by Coey [26] for the purpose of the reduction of $M_S$ in the maghemite ferrimagnetic particles.

The Weizsäcker model yields the equation for the magnetic moment $\mu(N) = a + b/N^{1/3}$ [8]. To test this hypothesis we apply the Monte Carlo method. This approach enables to obtain rather accurate quantitative results and is widely used in nanophysics now. In particular, a number of calculations for magnetically uniform clusters have been carried out [27- 31].

The investigation of the surface effects in the magnetic NPs FePt was presented by Labaye *et al.* [32]; the authors considered the effect of the surface anisotropy on an isolated single-domain spherical NP using atomic Monte Carlo simulation of the low-temperature spin ordering. The analogous behavior was found in the work [33] where the effect of surface anisotropy upon the magnetic structure of ferrimagnetic maghemite NPs was studied by applying the three-dimensional classical Heisenberg–Hamiltonian and the Monte Carlo methods. The calculations for the spherical particle where surface magnetic moments are different from bulk ones are performed.

The results reveal throttle structure with increasing surface anisotropy, as well as a significant decrease of the Curie temperature of the NPs as compared to that of the bulk maghemite. This difference can be due to the effects either of the coverage of magnetite NPs or of the surface anisotropy (see also the discussion of the microscopic magnetite model with the account of various anisotropic contributions [29]). From the Hamiltonian model (as a part of the classical Heisenberg one) it follows:

$$H = -\frac{J}{2}\sum_{\langle i,j \rangle} \mathbf{S}_i \cdot \mathbf{S}_j \qquad (7)$$

where $\langle i, j \rangle$ denotes nearest-neighbor sites of the spherical particle, and $\mathbf{S}_i$ is the atomic magnetic moment. Magnetic moments of atoms located in the bulk of the particles are normalized by unity, $|\mathbf{S}_i| = 1$, whereas on the surface $|\mathbf{S}_i| = s$, where $s$ is allowed to be different from 1 (the possible difference in exchange parameters for bulk and surface bonds can be taken into account by rescaling $s$). The spherical particle with the radius $R$ can be defined as a region of simple cubic lattice by inequality: $|\mathbf{r}_i| < R$, where $\mathbf{r}_i$ is lattices sites. Then surface layer is defined by inequality: $R - a < |\mathbf{r}_i| < R$, where $a$ is the lattice constant. However, it was found that these definitions lead to strong "geometrical" oscillations of the number of surface atoms $N_s$, which make the subsequent interpretation of the Monte Carlo results difficult. To reduce $N_s$ fluctuations, the definitions were modified by introducing a weak randomness. Specifically, the substitution $R \rightarrow R + x$ is made, where $x$ is a normally distributed variable with zero mean value and with standard deviation of $\sigma = 0.2$.

The Monte Carlo simulation was performed by modified heat bath algorithm [34] using ALPS library [35]. The $N$-dependence of full magnetic moment at the temperature $T = 0.3\ J$ which is lower than the Curie temperatures for all $N$ under consideration is shown in Figure 8. It is clear from the figure that if $s = 1$ the finite-size effects alone cannot provide a visible $N$-dependence for $N > 1000$. With the increase of $N$ the full magnetic moment decreases for $s > 1$ and increases for $s < 1$. These dependences are in accordance with data presented in figures 1 and 2. Also, a quantitative correlation with the Weizsäcker model [8] is observed.

As it shown in figure 9, the dependence of magnetic moment at low temperatures is linear as well as in infinite classical system (in the quantum case this behavior is modified, but not reduced to the usual $T^{3/2}$ Bloch law [36, 37]). However, the coefficient characterizing this dependence is considerably changes with $s$.

Note that for the finite systems the temperature dependence of magnetic moment can be modified in accordance with the discreteness of the magnon spectrum (resulting, in particular, in the energy gap) and the lowering of the mean value of nearest neighbors [36]. As demonstrated by the performed Monte Carlo calculations for both open and periodic boundary conditions, the latter effect predominates. The dependence $\mu(T)$ is also changed with corresponding changes in the surface anisotropy, as show the Monte Carlo results [38].

It should be stressed that the change in the temperature of the magnetic moment of finite NPs can be substantially different from that in the predictions of the spin wave theory (SWT), so that the real $T_C$ may be considerably lower than that given by SWT. The finding seems to be similar to that for the infinite layered magnets where the self-consistent spin wave theory (SSWT) yields the decrease of $T_C$ by about 30-50 percents [39-40]. (At the same time, SSWT still overestimates $T_C$ in comparison with the Monte Carlo calculations, see, e.g., Fig.5 of Ref.41.) Further decrease of $T_C$ may be obtained in the field-theoretical approaches which are especially efficient in the case of low $T_C$ [40].

The Curie temperature $T_C$ of the observed finite system (see figure 10) was determined as a temperature where fluctuations of the particle magnetic moment are the strongest. As a measure of these fluctuations, the quantity: $\langle \mu^2 \rangle - \langle |\mu| \rangle^2$ is used, where $\mu$ is the full magnetic moment of the particle. The $N$-dependence of the Curie temperature can be rather strong, with $T_C$ decrease for $s \leq 1$ and increase for $s > 1$. The $N$-dependence of $T_C$ is found to be stronger than that of magnetic moment.

Moreover, it was not expected that the $N$-dependence of the Curie temperature is not monotonous for $s > 1$. This non-monotonous behavior can result from the competition of the two factors: a reduced coordination number of the surface atoms, which favors disorder and the increase of individual magnetic moments of the surface atoms, which favors ordering. Although the maximal value of $N$ achieved in the simulations does not cover the whole investigated $N$-region, the calculations qualitatively reproduce the experimental behavior of the Curie temperature and a magnetic moment. Moreover, received Monte Carlo results provide some quantitative estimation, unlike those received using the phenomenological Weizsäcker model approach [8].

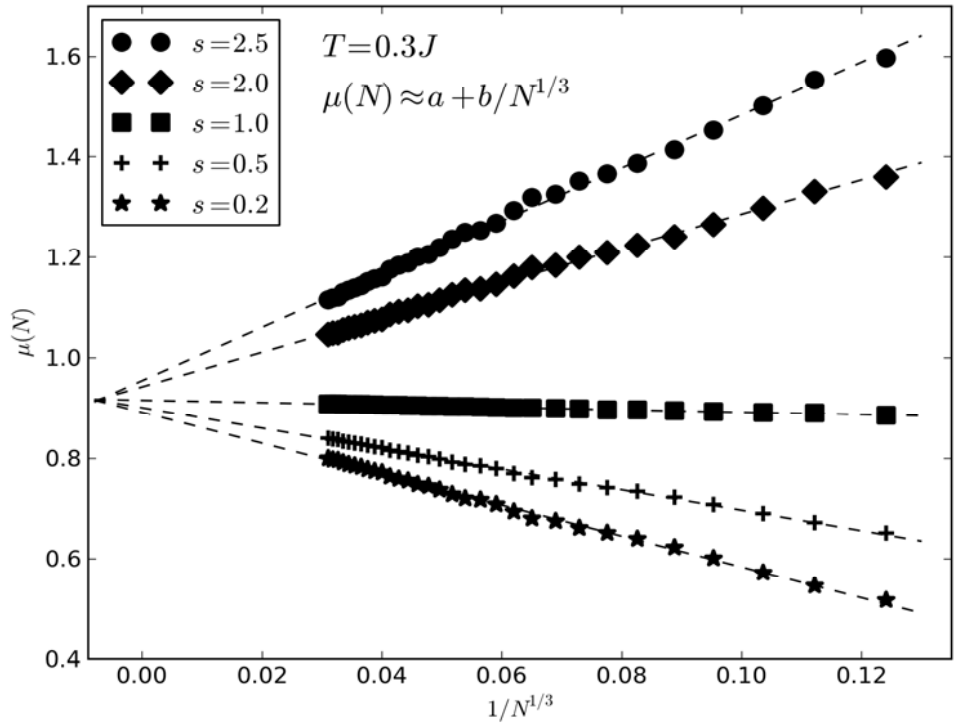

**Figure 8** Dependence of the full magnetic moment of the NP (normalized to one atom) on $1/N^{1/3}$ ($N$ is the number of atoms) for the spherical particle with different surface moments ($s$ values) and temperature $T = 0.3\ J$. Dashed lines are linear fits.

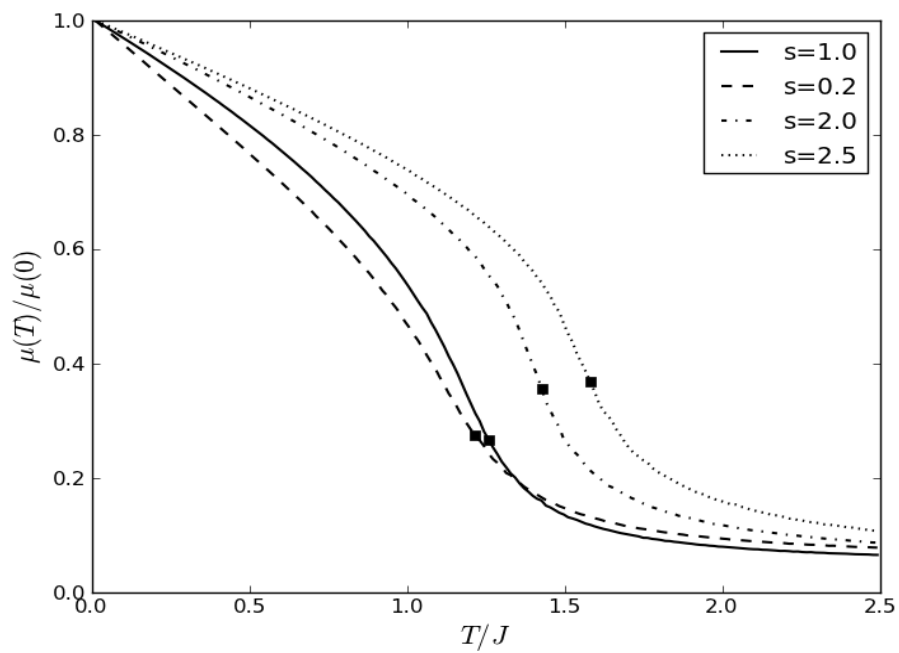

**Figure 9.** Typical temperature dependences of magnetic moment of NP containing *N*=515 atoms with different surface moments (*s* values*)*.

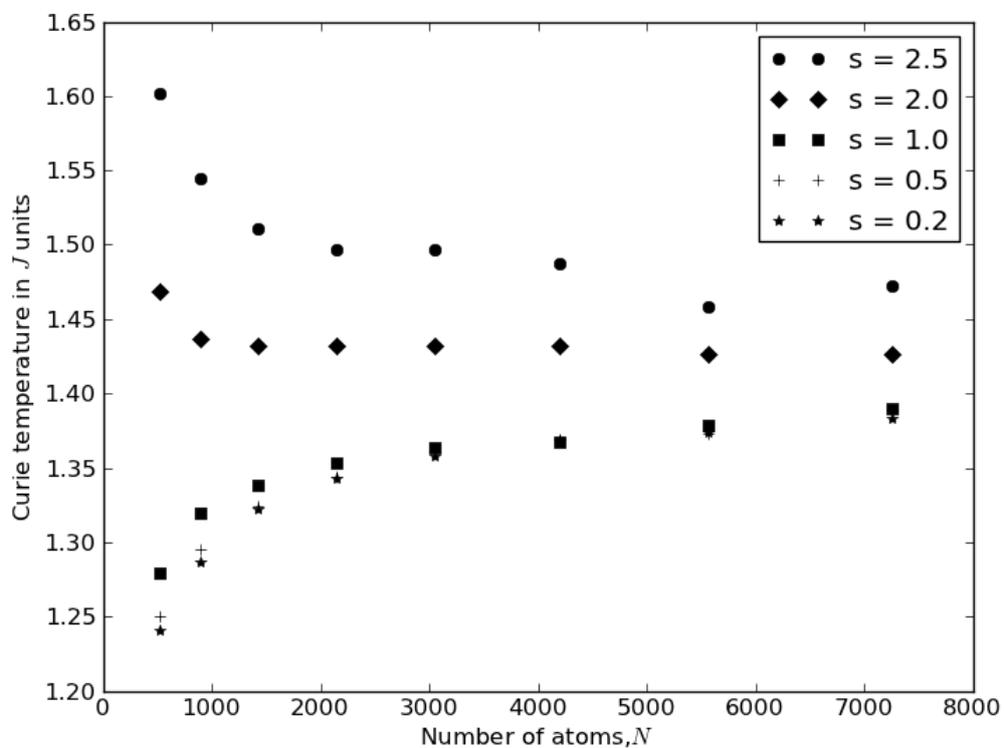

**Figure 10.** Dependence of the Curie temperature on the number of atoms in the spherical particle for different *s*.

## 5. Conclusions

The impact of the size of magnetite cluster on its magnetic properties (magnetic moment, the Curie temperature, blocking temperature, etc.) has been studied. The application of the magnetic separation and centrifugation methods for the aqueous biocompatible iron oxide nanoparticles (NPs) allow producing "monodisperse" fraction of NPs with different diameters (4÷22 nm) which have been examined by laser correlation spectroscopy, transmission electron microscopy, and X-ray diffractometry. For the first time, the method of the laser correlation spectroscopy has been applied in real time for the separation and control of the NP sizes in aqueous suspensions. Both intact NPs and those covered with a bioresorbable layer have been investigated. The results obtained are interpreted on the basis of Monte Carlo simulations for the microscopic Heisenberg model. All the results obtained are in a qualitative correlation.

It has been demonstrated experimentally and by theoretical modeling that magnetic properties of magnetite NPs were determined not only by their sizes, but also by the their surface spin states, while both growing and falling relationship of the magnetic moment (per $Fe_3O_4$ formula unit) being possible, depending on the number of magnetic atoms in the NP. The combination of the informative methods of magnetic diagnostics with phenomenological and microscopic models for NPs allows obtaining the relationship of the specific magnetic moment of NPs on the number of its constituent magnetic formula units, as well as it defines the contributions from the bulk and surface states to magnetization. For the first time, both experimental and theoretical temperature relationship of the magnetic properties of the magnetite NPs with different sizes, and of the bulk $Fe_3O_4$ sample have been obtained and compared.

Both intact NPs and those covered with a bioresorbable layer have been investigated. The type of the NP coating affects a number of physical characteristics of the studied NPs, in particular, their magnetic properties. Using the Langevin formula analysis of the magnetization curves M(H) for the suspension of superparamagnetic NPs, the magnetic moment per one NP could be calculated. The relationship of the magnetic moment per one formula unit for the magnetite NPs with dextran shell has decaying character with the increase of the number of formula units N, while the magnetite NPs in the polymer matrix demonstrate an opposite effect.

The change in the NP size controlled by DLS method causes the change in the Curie temperature and in the magnetic moment of NPs. With the decrease of the nanoparticle diameter, the Curie temperature decreased sharply. The relationship of the size of the magnetic moment on the size of magnetite NPs defines the value of the Curie temperature which is substantially lower than for the bulk samples. Unlike "ideal" NPs with identical bulk and surface individual magnetic moments, the calculated relationship of the magnetic properties in the core-shell model is found to be quite strong. As regard for the potential applications of the findings, they could be used in medicine, biotechnology, materials science and in forensic applications [42, 43].

ACKNOWLEDGEMENTS

This work is, in part, supported by the Program of fundamental research of RAS Physical Division, project No. 15-8-2-9 and by the "Dynasty" foundation. The numerical calculation were performed using the «Uran» supercomputer of Institute of Mathematics and Mechanics (the Urals Branch of the RAS).

REFERENCES

1. Z. M. Li, J. Shen, X. M. Sun, and Y. J. Wangel, Laser Phys. Lett. **10** 095701(2013)
2. S. Tanev, V. V. Tuchin, and P. Paddon, Laser Phys. Lett. **3**, 594(2006)
3. V. K. Pustovalov, V. A. Babenko, Laser Phys. Lett. **1**, 516 (2004)
4. *Nanoscale Materials in Chemistry* ed. K.J. Klabunde (New York: Wiley, 2001)
5. *Nanoparticles. From Theory to Application* ed. G. Schmid (Weinheim: Willey-VCH, 2004)
6. X. Battle and A. Labarta, *J. Phys D.: Appl. Phys.* **35** (2002) R15
7. V.N. Nikiforov, E.Yu. Filinova. In: *Biomedical application of magnetic nanoparticles.* Wiley-VCH

Verlag GmbH, 2009, p. 393-455
8. V.N. Nikiforov, B.L. Oxengendler, N.N. Turaeva, A.V. Nikiforov, and V.G. Sredin, *J. Phys.: Conf. Ser.* **291** (2011) 012009; V.N. Nikiforov et al, arXiv:1206.6985
9. P. Suzdalev, *Physical Chemistry of Nanoclusters, Nanostructures and Nanomaterials* (Moscow:Com.Kniga; 2006, *in Russian*)
10. Ch . Pool and F. Owens, *Introduction to Nanotechnology* (New York: Wiley–Interscience, 2007)
11. G. Kellermann and A. Graievich, *Phys. Rev. B* **65** (2002) 134204
12. G. Kellermann and A Graievich, *Phys. Rev. B* **78** (2008) 054106
13. J. Pakarinen, M. Backman, F. Djurabekova, and K. Nordlund, *Phys. Rev. B* **79** (2009) 085426
14. S.P. Gubin, J.A. Koksharov, G.B. Homutov, and G.J. Jurkov, *Uspekhi Khimii* **74** (2005) 539 [*Russian Chemical Reviews* **74** (2005) 489]
15. N.A. Brusentsov, V.D. Kuznetsov, T.N. Brusentsova, et al. *J. Mag. Magn. Mat.* **272-276**, (2004) 2350.
16. A.I. Autenshlyus, N.A. Brusentsov, A. Lockshin, *J. Magn. Magn. Mater.* **122** (1993) 360.
17. T.N. Brusentsova, N.A. Brusentsov, V.D. Kuznetsov, and V.N. Nikiforov, *Mag.Magn.Mat.* **293** (2005) 298
18. G.B. Khomutov, In: *Nanomaterials for Application in Medicine and Biology*, Ed. M. Giersig and G.B. Khomutov, Springer, 2008, p.p.39-58.
19. A.A. Novakova, V.Y. Lanchinskaya, A.V. Volkov, T.S. Gendler, T.Y. Kiseleva, M.A. Moskvina, S.B. Zezin, *J. Mag. Magn. Mat.* **258** (2003) 354
20. V.N. Nikiforov, Yu.A. Koksharov, S.N. Polyakov, A.P. Malakho, A.V. Volkov, M.A. Moskvina, G.B. Khomutov, V.Yu.Irkhin, *J. Alloys and Compounds.* **58** (2013) 569
21. V.N. Nikiforov, V.D. Kuznetsov, Yu.D. Nechipurenko, V.I. Salyanov, Yu.M. Evdokimov, *JETP Letters* **81** (2005) 327; V.N. Nikiforov, V.G. Sredin, A.V. Nikiforov et al. *Nanotechnology International Forum.6-8 Oct.* 254 (2009) 2 (*in Russian*)
22. K. Brymora, F. Calvayrac, arXiv:cond-mat/1205.1842 (2012)
23. H.C. Siegman, *J. Phys.: Condens. Matter.* **4** *(1992)* 8395
24. J.L. Dormann, D. Fiorani, D. Tronc *Advan. Chem. Phys.* **98** (1997) 283
25. A.E. Berkowitz, W.J. Shuele, P.J. Flanders, *J. Appl. Phys.* **39** (1968) 1261
26. J.M.D. Coey, *Phys. Rev. Lett.* **27** (1971) 1140
27. D.A. Dimitrov and G.M. Wysin, *Phys. Rev. B* **50** (1994) 3077
28. V.S. Leite and W. Figueiredo, *Braz. J. Phys.* **36** (2006) 652
29. J. Mazo-Zuluaga, J. Restrepo, J. Mejia-Lopez, *J. Appl. Phys.* **103** (2008) 113906
30. D. Kechrakos, *Handbook of Nanophysics* vol 3, ed K Sattler (Taylor & Francis, 2010)
31. E.Y. Vedmedenko, N. Mikuszeit, N. Stapelfeldt, R. Wieser, M. Potthoff, A.I. Lichtenstein, and R.Wiesendanger, *Eur. Phys. J.* B **80** (2011) 331
32. Y. Labaye, O. Crisan, L. Berger, J.M. Greneche, J.M.D. Coey, *J. Appl. Phys.* **91** (2002) 8715
33. J. Restrepo, Y. Labaye, J.M. Greneche, *Physica B* **384** (2006) 221
34. D. Loison, C.L. Qin, K.D. Schotte, and X.F. Jin, *Eur. Phys. J. B* **41** (2004) 395
35. B. Bauer et al. (ALPS collaboration) *J. Stat. Mech.* **P05001** (2011) 1742
36. P.V. Hendriksen, S. Linderoth, P.A. Lindgard, *Phys.Rev.B* **48** (1993) 7259
37. R.H. Kodama, *J.Mag.Magn.Mat.* **200** (1999) 359
38. K.N. Trohidou, *Monte Carlo Studies of Surface and Interface Effects in Magnetic Nanoparticles, chapter in Surface Effects in Magnetic Nanoparticles,* ed. D. Fiorani, (Springer, 2005)
39. V.Yu. Irkhin, A.A. Katanin, M.I. Katsnelson, *Fiz. Met. Metalloved.* **79**(1) (1995) 65 [*Phys. Met.Metallography* **79** (1995) 42].
40. A.A. Katanin, V.Yu. Irkhin, *Physics-Uspekhi* **50** (2007) 613
41. A.N. Ignatenko, A.A. Katanin, and V.Yu. Irkhin, *JETP Letters* **97** (2013) 209
42. R. Wiltschko, W. Wiltschko, *Magnetic Orientation in Animals* (Berlin: Springer, 1995)
43. J.L. Kirschvink, *Bioelectromagnetics* **10** (1989) 239